\documentstyle[12pt]{article}    
\begin{document}
\begin{center}
   
        {\bf    Finite-Temperature Casimir Effect on the Radius
Stabilization of Noncommutative Torus}

\vspace{1cm}

                      Wung-Hong Huang\\
                       Department of Physics\\
                       National Cheng Kung University\\
                       Tainan,70101,Taiwan\\

\end{center}
\vspace{2cm}

   The one-loop correction to the spectrum of Kaluza-Klein system for the
$\phi^3$ model on $R^{1,d}\times (T_\theta^2)^L$  is evaluated in the high
temperature limit, where the $1+d$ dimensions are the ordinary flat
Minkowski spacetimes and the $L$ extra  two-dimensional tori are chosen to
be the  noncommutative torus with noncommutativity $\theta$.  The
corrections to the Kaluza-Klein mass formula are evaluated and used to
compute the Casimir energy with the help of the Schwinger perturbative
formula in the zeta-function regularization method.    The results show
that the one-loop Casimir energy is independent of the radius of torus if
$L=1$.    However, when $L>1$ the Casimir energy could give repulsive force
to stabilize the extra noncommutative torus if  $d-L$ is a non-negative
even integral.   This therefore suggests a possible stabilization mechanism
of extra radius in high temperature, when the extra spaces are
noncommutative.  
 
\vspace{3cm}

\begin{flushleft}
   
E-mail:  whhwung@mail.ncku.edu.tw

\end{flushleft}
\newpage

\section  {Introduction}

The Casimir effect originally suggested in 1948 [1,2] is regarded as the
contribution of a non-trivial geometry on the vacuum fluctuation of quantum
electromagnetic fields.  The corresponding change in the vacuum energy can
produce a vacuum pressure to attract the two perfectly conducting parallel
plates.   This attractive force is experimentally confirmed by Sparnaay in
1958  and recently more precise measurement have been provided [3].

Appelquist and Chodos [4] in 1983 suggested that the vacuum fluctuation of
higher dimensional gravitational field may contribute an attractive Casimir
force to push the size of the extra spaces in the Kaluza-Klein unified
theory [5] to the Planck scale.   Near the Planck scale, it is generally
believed that the nonperturbative quantum gravity can stabilize the size of
the extra spaces [5].   The superstring theory, as an only candidate of
unified theory including the quantum gravity,  is also a higher dimensional
theory [6].   Therefore it is reasonable to conjecture that the string
theory could itself provide a mechanism to stabilize the extra compact
space therein.

In recent investigations, an available scale which may be used to stabilize
the extra compact is the noncommutativity $\theta_{ij}$ revealed in the
string/M theories [7-9].    Initially,  Connes, Douglas and Schwarz [7] 
had shown that the supersymmetric gauge theory on noncommutative torus is
naturally related to the compactification of Matrix theory [9].   More
recently,  it is known that the dynamics of a D-brane [10] in the presence
of  a B-field can, in certain limits, be described by the noncommutative
field theories [8].   

   Historically,  it is a hope that, introducing a parameter to deform the
geometry in the small spacetime would be possible to cure the quantum-field
divergences, especially in the gravity theory [11].    The later works
[12], however, proved that the noncommutative field theory exhibits the
same divergence as the commutative one.  But the field theory with
noncommutativity could show some interesting properties and still deserves
furthermore investigated. 

A distinct characteristic of the  noncommutative field theories, found by
Minwalla, Raamsdonk and Seiberg [12], is the mixing of ultraviolet (UV) and
infrared (IR) divergences reminiscent of the UV/IR connection of the string
theory.  In a recent paper [13] we also found that the noncommutativity
does not affect one-loop effective potential of the scalar field theory.   
However, it can become dominant in the two-loop level and have an
inclination to induce the spontaneously symmetry breaking if it is not
broken in the tree level, and have an inclination to restore the symmetry
breaking if it has been broken in the tree level.  

In the paper [14] Nam used the Kaluza-Klein mass formula, which was
evaluated by Gomis, Mehen and Wise [15], to compute the one loop Casimir
energy of an interacting scalar field in a compact noncommutative space of
$R^{1,d}\times T^2_\theta$, where $1+d$ dimensions are the ordinary flat
Minkowski space and the extra two dimensions are noncommutative torus with
noncommutativity $\theta$.   It is the first literature which tries to use
the noncommutativity as a  minimum scale to protect the collapse of the
extra spaces.  

In the paper [16] we followed the method of [15] to evaluate the one-loop
correction to the spectrum of Kaluza-Klein system for the $\phi^3$ model on
$R^{1,d}\times (T_\theta^2)^L$, where the extra dimensions are the L
two-dimensional noncommutative tori .   We used the correct Kaluza-Klein
mass spectrum [16] to compute the Casimir energy.    It then shows that
when $L>1$ the Casimir energy due to the noncommutativity could give
repulsive force to stabilize the extra noncommutative tori in the cases of
$d = 4n - 2$, with $n$ a positive integral.   This therefore suggests a
possible stabilization mechanism for a scenario in superstring theory,
where some of the extra dimensions are noncommutative. 

In this paper we will  investigate the previous problem, while in the
non-zero temperature.   Note that the Casimir energy for the system in
ordinary commutative spacetime is always found to attract the two perfectly
conducting parallel plates for all temperature [1,2].   So, it is
interesting to see how the noncommutativity will affect the behaviors of
finite-temperature Casimir energy.

Note also that the Casimir effect is null in the supersymmetry system, as
the
contribution from boson field is just canceled by that from fermion field
[17].   The no-go theorem of supersymmetry breaking [17] tells us that the
supersymmetry could not be radiatively broken,  even if the gauge symmetry
has been broken in the tree level [18].  However, some mechanisms shall be
proposed to break the supersymmetry in superstring theory to describe the
physical phenomena [6,17].   Thus the remaining non-zero Casimir effect may
be used  to render the compact space stable.    An interesting mechanism
which could break the supersymmetry is the temperature effect [19], a
natural consequence of different statistics for bosons and fermions.   This
is the scenario in the early epoch of the universe and is related to the
present paper.

In section II, we extend the works of Gomis, Mehen and Wise [15] to the
non-zero temperature.   We evaluate the one-loop correction to the spectrum
of Kaluza-Klein system in the high temperature limit.  The correction to
the Kaluza-Klein mass formula has the additional term which resembles that
of the winding states in the string theory [20], likes as the property
found in the zero temperature system [15].   In section III, the obtained
spectrum is used to compute the Casimir energy with the help of the
Schwinger perturbative formula in the zeta-function regularization method
[21].    The result is used to analyze the effect of the high-temperature
Casimir energy on the radius stabilization.   We find that the one-loop
Casimir energy is independent of the radius of torus if $L=1$.    So, in
this case, there is no repulsive force and we have to consider the Casimir
energy to the higher order, which is beyond the scope of this paper. 
However, when $L>1$ the Casimir energy could give repulsive force to
stabilize the extra   noncommutative torus if  $d-L$ is a non-negative even
integral.   This therefore suggests a possible stabilization mechanism for
a high-temperature scenario in the Kaluza-Klein theory, where some of the
extra dimensions are noncommutative.   The conclusion is presented in the 
last section.


\section {Kaluza-Klein Spectrum for $\phi^3$ on $R^{1,d} \times
(T^2_\theta)^L$ : Finite Temperature}

\subsection{Model}

It is known that the algebra of functions on a noncommutative spacetime can
be viewed as an algebra of ordinary functions on the usual spacetime with
the product deformed to noncommutative Moyal product [12].   Therefore, the
scalar $\phi^3$ theory in $R^{1,d} \times (T^2_\theta)^L$ spacetime can be
described by the following action:

$$ S = \int d^{1+d} x ~ d^{2L}y \left( {1\over 2} (\partial \phi)^2
-{1\over 2} m^2 \phi^2 - { \lambda \over 3!} \phi \star \phi \star
\phi\right), \eqno{(2.1)}$$ 
\\
in which the $\star$ operator is the Moyal product generally defined by
[12]

$$f(x) \star  g(x) = e^{+{i\over 2} \theta^{\mu\nu} {\partial\over \partial
y^\mu} 
{\partial\over \partial z^ \nu} } f(y) g(z) |_{y,z\rightarrow x}.  
\eqno{(2.2)} $$
where $\theta_{\mu\nu}$ is a real, antisymmetric matrix which represents
the noncommutativity of the spacetime, i.e., $ [x^\mu,x^\nu] = i \theta
^{\mu \nu}$. 
In Eq.(2.1) the coordinates $x^0,x^1,..., x^d$ represent the commutative
four dimensional Minkowski spacetime.  The $2L$ extra dimensions are taken
to be the $L$ noncommutative 2-tori $T^2_\theta$ whose noncommutative
coordinates are described by
$$ [y^1,y^2] =[y^3,y^4] = ... = [y^{2L-1},y^{2L}] = i \theta.  
\eqno{(2.3)}$$ 
When $L=1$, this coordinate can be realized in string theory by wrapping a
five-brane on a two-torus $T^2$ with a constant $B$-field along the torus. 
The low energy effective four dimensional theory resulting from
compactification on a noncommutative space is local and Lorentz invariant,
hence it can be relevant phenomenologically [15]. 

\subsection{One-Loop Calculation}

   At zero temperature, the momentum in the $1+d$ Minkowski spacetime,
denoted as $p$, is a continuous variable.   However, the momenta along the
tori are quantized as
$\vec k \to \vec k /R$, where $\vec k = (k_1 ..., k_{2L})$ are the
integrals.  Therefore, using the Feynman rule [15], which  includes the
propagator $i \Delta (p, \vec n) = \frac {i (1- \delta_{\vec n ,0})}{p^2 -
{\vec n}^2 - m^2}$ for the field with momentum $(p, \vec n)$, and vertex
function $V(\vec k,\vec n,\vec m) = - i \lambda ~ cos( \frac {\theta}{2R^2}
\vec n \wedge \vec k) \delta_{\vec k +\vec n +\vec m,0}$ for the three legs
with  incoming momenta $\vec k,\vec n$ and $\vec m$ respectively, the
one-loop contribution to the two point functions is  

$$ {\lambda^2 \over 4} {1\over (2 \pi R)^{2L}} \sum_{\vec{k}} \int
{d^{1+d} l
 \over (2 \pi)^{1+d}} {\cos^2(\theta ~ \vec{n} \wedge \vec{k}/(2 R^2) )(1-
\delta_{\vec{k},0}) (1-\delta_{\vec{n}+\vec{k},0})\over (l^2 -\vec{k}^2/R^2
- m^2) ((l+p)^2 - (\vec{n}+\vec{k})^2/R^2 - m^2)}  $$

$$={\lambda^2 \over 4} {1\over (2 \pi R)^{2L}} \sum_{\vec{k}} \int {d^{1+d}
l
\over (2 \pi)^{1+d}} {1-2\, \delta_{\vec{k},0} -
2\delta_{\vec{n}+\vec{k},0}+\cos( \theta ~ \vec{n} \wedge \vec{k}/R^2) 
\over (l^2 -\vec{k}^2/R^2 - m^2) ((l+p)^2 -(\vec{n} +\vec{k})^2/R^2 - m^2)
}.    \eqno{(2.4)}$$
\\
in which  $\vec{n} \wedge \vec{k} \equiv (n_1 k_2 - n_2 k_1) +(n_3 k_4 -
n_4 k_3) ...+ (n_{2L-1} k_{2L} - n_{2L} k_{2L-1})$.   Note that $l$ denotes
the loop momenta along the noncompact directions, while $\vec{k}$ is loop
momenta along compact directions. Similarly, $p\, (\vec{n})$ is the
external momenta along the noncompact (compact) direction.   The derivation
of Eq.(2.4) has used the half angle formula for the cosine and the property
that $\vec{n} \wedge \vec{n} = 0$, as detailed by Gomis, Mehen and Wise
[15].

The first term, second term and third term in Eq.(2.4) are all divergent
[15].   They are irrelevant to our investigation and will not be discussed
furthermore.   The last term contains a oscillatory factor $\cos (\theta \,
\vec{n} \wedge \vec{k}/R^2)$ which makes the non-planar correction term to
be ultraviolet finite [15].  

Let us turn to the system at non-zero temperature.   It is known that the
divergences at non-zero temperature are the same as those at zero
temperature.   Therefore, to find the one-loop correction of the
Kaluza-Klein spectrum at finite temperature we only need to evaluate the
last term in Eq.(2.4).  

At temperature $T ~ (=1/ \beta)$ the non-planar correction to the one-loop
self energy is the last term in the Eq.(2.4) after substituting the
continuous variable $l_0$ by $\frac {2\pi l_0}{\beta}$, where the new
variable $l_0$ is an integral.  Therefore the one-loop self energy is 

$$\Sigma(p,\vec n) ={\lambda^2 \over 4} {1\over (2 \pi R)^{2L}} {1 \over
\beta}\sum_{\vec{k}} \sum_{l_0}\int {d^d \vec{l} \over (2 \pi)^d} \times
\hspace{8cm} $$
$${\cos (\theta ~ \vec{n} \wedge \vec{k}/R^2) \over [(\frac {2\pi
l_0}{\beta})^2 -  {\vec l}^2 -\vec{k}^2/R^2 - m^2] [(\frac
{2\pi}{\beta}(l_0 +p_0))^2 - (\vec l+\vec p)^2 -(\vec{n} +\vec{k})^2/R^2 -
m^2] } ~.    \eqno{(2.5)}$$
\\
To proceed, after introducing the Feynman parameter $x$ we can perform the
integration over the momentum $\vec l$, then expressing the result in terms
of the Schwinger parameter $\alpha$ the equation becomes  

$$\Sigma(p,\vec n) ={\lambda^2 \over 4} {1\over (2 \pi R)^{2L}} {1 \over
\beta} {1 \over (4\pi)^{2/d}}\sum_{\vec{k}} \sum_{l_0} \int_0^1 dx 
\int_0^\infty\ d\alpha \alpha ^{1-d/2} \times \hspace{5cm} $$
$$ ~ exp \left\{-\alpha \left[m^2+x(1-x) {\vec p}^2 + {{\vec k}^2 + x (2
\vec k \cdot \vec n + {\vec n} ^2)  \over {R^2}} + \frac{l_0^2 + x (2 l_0
p_0 + p_0 ^2)} {(1/ 2\pi \beta)^2}\right]\right\} \times $$
$$ {1\over 2}\left \{ \prod _{\scriptstyle i=1}^{2L-1}  exp[i
\frac{\theta}{R^2}( n_i k_{i+1}-  n_{i+1} k_i) ] +  \prod _{\scriptstyle
i=1}^{2L-1}  exp[- i \frac{\theta}{R^2}( n_i k_{i+1}-  n_{i+1} k_i) ]
\right \} .$$
\\
We can now perform the sum over $\vec{k}$ in the above equation by using
the definition of the Jacobi theta function 

$$ \vartheta(\nu, \tau) = \sum_{n=-\infty}^{\infty} \exp(\pi i n^2 \tau +2
i\pi n \nu).    \eqno{(2.6)}$$ 

\noindent
Then using the property of modular transformation [2,14]
  
$$\vartheta(\nu,\tau)=(-i \tau)^{-1/2}\exp( -\pi i \nu^2/\tau)
\vartheta(\nu/\tau, -1/\tau).    \eqno{(2.7)}$$

\noindent
we can express the result as

$$\Sigma(p,\vec n) = - {\lambda^2 \over 4} {1\over (2 \pi R)^{2L}} {1 \over
\beta} {1 \over (4\pi)^{2/d}} \sum_{l_0} \int_0^1 dx  \int_0^\infty\
d\alpha \alpha ^{1-d/2} ~ ({\alpha \over \pi R^2})^{-L}  ({\beta^2 \over
4\pi\alpha})^{1/2} ~ \times \hspace{2cm} $$
$$ ~ exp \left \{-\alpha \{m^2+x(1-x) [~ ({\beta p_0 \over 2\pi})^2 ~ +
{\vec p}^2 + {{\vec n} ^2 \over {R^2}}] ~ \} - {1\over \alpha}{\theta ^2
\vec n^2\over 4 R^2} \right\} exp \left \{- {\beta^2 \over 4 \alpha}~
\l_0^2 + i 2\pi x p_0 l_0\right\} \times $$
$$ \prod _{\scriptstyle j=1}^{2L-1} \vartheta (x n_j + i {n_{j+1}\over
2\alpha},   i {\pi R^2\over \alpha}) ~ \vartheta (x n_{j+1} - i {n_j\over
2\alpha}, i {\pi R^2\over \alpha}) .$$
\\
From the above equation we see that the ultraviolet divergent contribution
comes from $\alpha \to 0$ region.   Thus the leading contribution will come
from the region  $\alpha \to 0$ [15] and  we can set $\vartheta =1$ in the
above equation. The leading correction to the spectrum of Kaluza-Klein
system becomes

$$\Sigma(p,\vec n) \approx - {\lambda^2 \over 4} {1\over (4
\pi)^{L+{d+1\over 2}}}  \sum_{l_0}  \int_0^1 dx \int_0^\infty\ d\alpha
\alpha ^{1-L-{1+d\over2}} ~ exp \left \{- {1\over \alpha}{\theta ^2 \vec
n^2\over 4 R^2} \right\} exp \left \{- {\beta^2 \over 4 \alpha}~ \l_0^2 + i
2\pi x p_0 l_0\right\} $$
$$= - {\lambda^2 \over 4} {1\over (4 \pi)^{L+{d+1\over 2}}}  \sum_{l_0} 
\int_0^1 dx \frac{\Gamma(L+{d-3\over2})}{[{\beta^2 l_0^2\over 4}+{\theta^2
\vec n^2\over 4 R^2}]^{L+{d-3\over2}}}~ exp( i 2\pi x p_0 l_0) .
\hspace{4cm}\eqno{(2.8)} $$
\\
This is the main result of this section.   

\subsection {Self Energy}
Let us first examine the above result at zero temperature.    In this case
$\beta \to\infty$ and only the mode $l_0=0 $ contribute the summation in
Eq.(2.8). Thus we can easily find that 

$$\Sigma(p,\vec n) = - {\lambda^2 \over 64} {\Gamma(L+{d-3\over2})\over 
\pi^{L+{d+1\over 2}}} \left ({R^2 \over  \theta^2 \vec n^2}\right
)^{L+{d-3\over2}} ,  ~~~~~ if ~~T = 0,  \eqno{(2.9)}$$
\\
which is that derived by us in [16] and reduces to that in [15] when $L=1$
and $d=3$.

From Eq.(2.8) we also see that if $p_0\ne 0$ (note that both of $p_0$ and
$l_0$ are  integral) then the integration over $x$ will becomes zero unless
$l_0 =0$. Thus we have another relation

$$\Sigma(p,\vec n) = - \lambda_p^2 ~ \left ({R^2 \over  \theta^2 \vec
n^2}\right )^{L+{d-3\over2}} ,  ~~~~ \lambda_p^2 \equiv  {\lambda^2 \over
64} {\Gamma(L+{d-3\over2})\over  \pi^{L+{d+1\over 2}}}  ,  ~~~~~ if ~~p_0
\ne 0, \eqno{(2.10)}$$
\\
When $p_0 =0$ then, as $\beta \to 0$ in the high-temperature limit, the
summation over $l_0$ in Eq.(2.8) could be replaced by an integration.   The
result  is 

$$\Sigma(p,\vec n) = - {\lambda_0^2 \over \beta} \left ({R^2 \over 
\theta^2 \vec n^2}\right )^{L+{d\over2}-2} , ~~~ \lambda_0^2 \equiv 
{\lambda^2 \over 64 } {\Gamma(L+{d\over2}-2)\over  \pi^{L+{d\over 2}}}  ,
~~~~if ~~ T >> 1, ~~ p_0 = 0. \eqno{(2.11)} $$
\\
The relation (2.9) have been used to evaluated the zero-temperature Casimir
energy in [16].  We will in next section use the relations (2.10) and
(2.11) to evaluated the Casimir energy in the high-temperature limit.

Note that from Eqs.(2.10) and (2.11) we see that when ${L+{d-3\over2}}=1$
in the case of  $p_0\ne 0$ or ${L+{d\over2}-2}=1$ in the case of $p_0=0$,
then the winding states like as those in the string theory will appear
[20].  The property had been found in the zero temperature system [15].  

\section {Casimir Energy at Finite Temperature}

\subsection {Partition Function, Zeta Function and Casimir Energy}

For the model (2.1) in thermal equilibrium at a finite temperature the
partition $Z$ for the system can be evaluated from the relation [21]

    $$ln Z = - {1 \over 2} \zeta_H^\prime (0) . \eqno{(3.1)} $$
\
in which the zeta function is defined by 
\
   $$\zeta_H (\nu) = {1\over \Gamma(\nu)}\int ds s^{\nu -1} \sum_{p_0} \sum
_{\vec n} \int d^d {\vec p} ~ e^{-s H},  \eqno{(3.2)}$$
\\
where $H = H_0 + \Sigma(p,\vec n), H_0 = ({2\pi p_0 \over \beta})^2 + {\vec
p}^2 + {\vec n ^2 \over R^2}$ and $\Sigma(p,\vec n)$ is defined by (2.10)
and (2.11).   The zeta function can be evaluated with the help of the
Schwinger perturbative formula [20]

$$\zeta_H (\nu) =\zeta_0 (\nu) + \zeta_1 (\nu),  \eqno{(3.3)}$$
\
in which 
$$\zeta_0 (\nu) = {1\over \Gamma(\nu)}\int ds ~ s^{\nu -1} \sum_{p_0} \sum
_{\vec n} \int d^d {\vec p} ~ e^{-s H_0 },  \eqno{(3.4)}$$

$$\zeta_1 (\nu) = {1\over \Gamma(\nu)}\int ds ~ s^{\nu -1} \sum_{p_0} \sum
_{\vec n} \int d^d {\vec p} ~ (- s ~ \Sigma(p,\vec n) ) ~ e^{-s H_0 }. 
\eqno{(3.5)}$$
\\
Using the usual formula [1,2] 

   $$E = {\partial \over \partial \beta }ln Z .   \eqno{(3.6)}$$ 
the Casimir energy is obtained.

\subsection {Calculation of Zeta Function}

Let us first evaluate the $\zeta_0(\nu)$.   We can first perform the
integration over the momentum $\vec p$ and then the variable $s$.  The
result is  

$$\zeta_0 (\nu) = {1\over \Gamma(\nu)}\int ds s^{\nu -1} \sum_{p_0} \sum
_{\vec n} \int d^d {\vec p} ~ exp \left [-s  \left (({2\pi p_0 \over
\beta})^2 + {\vec p}^2 + {\vec n ^2 \over R^2} \right ) \right ]
\hspace{2cm}$$

$$ = \pi^{d/2} ~ {\Gamma(\nu - {d \over 2})\over \Gamma(\nu)} \sum_{p_0}
\sum _{\vec n}  \left [({2\pi p_0 \over \beta})^2  + {\vec n ^2 \over R^2}
\right ]^{d/2 - \nu} \hspace{3.5cm}$$

$$ \approx \nu \left [ 2^{2L+d+1} \pi ^{-1/2} ~\beta ^{-2L-d}
~\Gamma({2L+d+1\over 2}) ~ \zeta (2L+d+1)  \right ] + O(\nu ^2), 
\eqno{(3.7)}$$
\\
where $\zeta(N)$ is the Riemann zeta function which is only divergent at
$N=1$.   Note that to obtain the final relation we have used the formula
A(12) in Ref. [2] to take the summations over  $p_0$ and $\vec n$ in the
high-temperature limit.   Substituting the above result into Eqs.(3.1) and
(3.6) we see that the zero-order Casimir force (i.e., without
noncommutativity) is attractive for arbitrary values of $L$ and $d$.

To calculate $\zeta_1 (\nu)$ we first consider  that contributes from $p_0
= 0$.   
From Eqs.(2.11) and (3.5) we have 

$$\zeta_1^{p_0=0} (\nu) = {1\over \Gamma(\nu)}\int_0^\infty ds ~ s^{\nu} 
\sum _{\vec n} \int d^d {\vec p} ~{\lambda_0^2\over\beta}  \left ({R^2
\over  \theta^2 \vec n^2}\right )^{L+{d\over2}-2} ~  exp \left [-s  ({\vec
p}^2 + {\vec n ^2 \over R^2}) \right ] , \hspace{1.5cm} $$

$$= \nu \left[ {1\over\beta} ~ {\lambda_0^2\over (\theta^2)^{L-2+d/2}} ~
{\pi^{d/2} ~ \Gamma(1+d/2)} ~ R^{2L-2} ~ \sum _{\vec n} \left ({1 \over 
\vec n^2 }\right )^{L-1} \right ] + O(\nu^2) .  \eqno{(3.8)}$$
\\
Substituting the above result into Eqs.(3.1) and (3.6) we see that the
contribution from $p_0 =0 $ in the first-order Casimir force is attractive
for arbitrary value $d$ if $L >1$.   When $L=1$ the Casimir energy is $R$
independent and there is no Casimir force. 

Let us turn to the contribution from $p_0 \ne 0$.  We first from the
Eqs.(2.10) and (3.5) perform the integration over $\vec p$. Then after
introducing a variable $x$ we can perform the integration over the variable
$s$ and then the variable $x$.   The result is
   
$$\zeta_1^{p_0\ne 0} (\nu) = {\lambda_p^2\over \Gamma(\nu)}\int_0^\infty ds
~ s^{\nu}  \sum _{\vec n}  \sum _{p_0\ne 0}\int d^d {\vec p} ~ \left ({R^2
\over  \theta^2 \vec n^2}\right )^{L+{{d-3}\over 2}} exp \left [-s  \left
(({2\pi p_0 \over \beta})^2 + {\vec p}^2 + {\vec n ^2 \over R^2} \right )
\right ] $$

$$={\lambda_p^2 \pi^{d/2}\over \Gamma(\nu) \Gamma(L + {{d-3}\over
2})}\int_0^\infty dx x^{L + {{d-5}\over 2}}\int_0^\infty ds ~ s^{L+\nu
-3/2}  \sum _{\vec n}  \sum _{p_0\ne 0} ~{\lambda_p^2\over\beta}  \left
({R^2 \over  \theta^2 \vec n^2}\right )^{L+{{d-3}\over 2}}  \times$$

$$ exp \left [-s  \left (({2\pi p_0 \over \beta})^2 + x {\vec n ^2 \over
R^2} \right ) \right ] $$

$$ =\nu \left[\lambda_\theta^2  ~ {1\over \Gamma(L-d-1/2)} ~ \beta^{2-3d} ~
R^{2L-2} \right] + O(\nu^2) , \hspace{3cm}\eqno{(3.9)}$$
where
$$\lambda_\theta^2 \equiv {\lambda^2\over {(\theta^2)}^{L + {d-3\over 2}}}
~ \pi^{L+5/2} ~ 2^{3d-6} ~ \Gamma({3d-1\over2}) ~ \Gamma(L + {d-3\over2}) ~
\sum_{\vec n}  \left ({1 \over  \vec n^2 }\right )^{2L+d-3} , 
\eqno{(3.10)}$$
\\
is positive.   Note that to obtain the above result we have use the
reflation formula [2]

$$\pi^{-{z\over2}} ~ \Gamma ({z\over2}) ~ \zeta (z) = \pi^{-{1-z\over 2}}
~\Gamma ({1-z\over 2}) ~ \zeta (1-z) ,        \eqno{(3.11)}$$
\\
to regularize the divergence in the original relation. 

Substituting the above result into Eqs.(3.1) and (3.6) we see that the
contribution from $p_0 \ne 0 $ in the first-order Casimir force may be
attractive or repulsive, which depend on the values of $L$ and $d$.   This
is because that the Gamma function $\Gamma(L-d-1/2)$ in (3.9) can become
negative if  $d-L$ is a non-negative even integral.   Not also that, as
that in the case of $p_0=0$, when $L=1$ the Casimir energy is $R$
independent and there is no Casimir force. 

\subsection{Compactification Radius}
Therefore, when $L>1$ and $d-L$ is a non-negative even integral, we can
substituting Eqs.(3.7), (3.8) and (3.9) into Eqs.(3.1) and (3.6) to find
the Casimir energy.   From the Casimir energy we can find the Casimir force
and find the compactification radius $R_T$.    In the high-temperature
limit we have the final result 

$$R_T =\left[- {(3d-2) (L-1) \pi^{1/2} ~ \lambda_\theta^2 ~ \beta^{2L+2-2d}
\over (2L+d) ~ 2^{2L+d+2} ~ \Gamma({2L+d+1\over 2}) ~ \Gamma(L-d-1/2)
\zeta(L+d+1)} \right]^{1/(2L+2)} .   \eqno{(3.12)}$$
\\
Thus we see that the Casimir energy could give repulsive force to stabilize
the extra noncommutative torus.  This therefore suggests a possible
stabilization mechanism for a high-temperature scenario in the Kaluza-Klein
theory, where some of the extra dimensions are noncommutative. 

\section{Conclusion}
The appearance of the parameter of noncommutativity in the context of
string theory [8] seems to be a good opportunity to provide a  minimum
scale to protect the collapse of the extra spaces.   This is because that
the commutation relation $[x^\mu,x^\nu] = i \theta^{\mu\nu}$ leads to the
spacetime-time uncertain relation $\delta x^\mu \delta x^\nu \geq {1\over
2} \theta^{\mu\nu}$.    However,  the noncommutativity will in general
break the unitarity [21] and Lorentz invariant [8]. Therefore, in this
paper we choose the spacetime in which noncommutativity only shows in the
extra space.   The result theory is unitary with the observable spaces
which are Lorentz invariant, as that consider by Gomis, Mehen and Wise in
ref. [15], and could be relevant phenomenologically

The investigation of Casimir effect in the non-zero temperature is
physically interesting,  as the temperature can break the supersymmetry and
Casimir energy contribution of boson field will not be canceled by that of
the fermion field.    The temperature in the early epoch of the universe is
very high and may relate to the present paper.

Our evaluations have shown that the Casimir force is attract if the extra
spaces are commutative.    However, when the L extra two-dimensional tori
become noncommutative, then it could contribute a repulsive force  to
protect the collapse of the extra spaces.  We have found that the one-loop
Casimir energy is independent of the radius of torus if $L=1$.    So, in
this case, there is no repulsive force and we have to consider the Casimir
energy to the higher order, which is beyond the scope of this paper. 
However, when $L>1$ the Casimir energy could give repulsive force to
stabilize the extra noncommutative torus if  $d-L$ is a non-negative even
integral.   We therefore have found a possible stabilization mechanism of
extra radius in high temperature, when the extra spaces are noncommutative.

Finally, we want to mention an interesting phenomena about our
investigations. 
At zero temperature, the previous work [16] found that when $L>1$ then the
Casimir energy due to the noncommutativity could give repulsive force to
stabilize the extra noncommutative tori in the cases of $d = 4n - 2$, with
$n$ a positive integral.   Comparing this with the present work at finite
temperatureone, we see that the noncommutative tori which will collapse at
zero temperature may, due to finite-temperature Casimir effect, have a
finite stable radius at high temperature.   On the other hand, that having
a stable radius at zero temperature may collapse at high temperature.   The
behavior depends on the values of $L$ and $d$.    Such a dramatic change,
however,  dose not show in the commutative space.  Therefore, the spaces
which are noncommutative may have phase transition at a finite temperature.
  The transition is due to the finite-temperature Casimir effect on a
noncommutative geometry.
\newpage
\begin{enumerate}
\item H.B.G. Casimir,{\it ``On the Attraction Between Two Perfectly
Conducting Plates''}, Proc. K. Ned. Akad. Wet, {\bf 51} (1948) 793;\\
G. Pluien, B. M\"uller, and W. Greiner, Phys. Rep. {\bf 134} (1986) 87;\\
V.M. Mostepanenko and N.N. Trunov, {\it ``The Casimir Effect and its
Applications''}, Oxford Univ. Press, 1997.
\item J. Ambjorn and S. Wolfram, Ann. Phys. {\bf 147} (1983) 1.
\item M. J. Sparnaay, Physica {\bf 24} (1958) 751; S. K. Lamoreaux, Phys.
Rev. Lett. {\bf 78} (1997) 5.
\item T. Appelquist and A. Chodos, Phys. Rev. Lett.{\bf 50} (1983) 141;
Phys. Rev. {\bf D28} (1983) 772. 
\item   Th. Kaluza, Situngsber. d. K. Preuss. Akad. d. Wissen.
z. Berlin, Phys.-Math. Klasse (1921) 966; O. Klein, Z. F. Physik 37 (1926)
895;\\
T. Appelquist, A. Chodos, and P.G.O. Freund,``{\it Modern Kaluza-Klein
Theories}'', Addison-Wesley, Menlo Park,  1987.
\item M. B. Green, J. H. Schwarz and E. Witten, {\it Superstring Theory},
Cambridge , 1987.
\item  A. Connes, M. R. Douglas and A. Schwarz,
  ``Noncommutative Geometry and Matrix Theory: Compactification on
  Tori'', hep-th/9711162, JHEP {\bf 02} (1998) 003; \\
 B.~Morariu and B.~Zumino, ``Super Yang-Mills on the
  Noncommutative Torus'', hep-th/9807198; \\
 C.~Hofman and E.~Verlinde, ``U-duality of Born-Infeld on the
Noncommutative Two-Torus'',  hep-th/9810116, JHEP {\bf 12} (1998) 010.
\item F. Ardalan, H. Arfaei and M. M. Seikh-Jabbari,  ``Noncommutative
Geometry from String and Branes'',  hep-th/9810072, JHEP {\bf 02} (1999)
016; ``Dirac Quantization of Open String and Noncommutativity in Branes'', 
hep-th/9906161, Nucl. Phys. {\bf  B 576} (2000) 578;M. M. Seikh-Jabbari and
A. Shirzad,``Boundary Conditions as Dirac Conditions'',  hep-th/9907055;\\
N.~Seiberg and E.~Witten,
``String Theory and Noncommutative Geometry'',  hep-th/9908142, 
JHEP {\bf 09} (1999) 032,\\
C. C. Chu and P. M. Ho, ``Noncommutative Open string and D-brane'', 
hep-th/9906192, Nucl. Phys. {\bf B 550} (1999) 151,\\
T. Lee, ``Canonical Quantization of Open String and Noncommutative
Geometry'',  hep-th/9911140, Phys. Rev {\bf D 62} (2000) 024022.
\item N. A. Obers and B. Pioline, ``U-Duality and M-Theory'' Phys. Rep.
{\bf 318} (1999) 113, hep-th/9809039. 
\item  J. Polchinski, {\it String Theory}, Cambridge University Press,
1998.
\item H. S. Snyder, Phys. Rev. {\bf 71} (1947) 38;  {\bf 72} (1947) 68;\\
A. Connes, {\it Noncommutative Geometry} Academic. Press, New York,
1994);\\
G. Landi G, {\it An introduction to noncommutative spaces and their
geometries}, Lecture Notes in Physics, Springer-Verlag , hep-th/97801078
;\\
J. V\'arilly, {\it An introduction to noncommutative geometry}, Summer
School "Noncommutative geometry and applications", Lisbon, September 1997 ,
physics/97090045. 
\item T. Filk, Phys. Lett. B 376 (1996) 53;\\
S.~Minwalla, M.~Van Raamsdonk and N.~Seiberg,
``Noncommutative Perturbative Dynamics'',
hep-th/9912072, JHEP {\bf 02} (2000) 020;\\
     C.~P.~Martin and D.~Sanchez-Ruiz,``The One-loop UV Divergent Structure
of U(1) Yang-Mills Theory on Noncommutative $R^4$'', hep-th/9903077, Phys.\
Rev.\ Lett.\  {\bf 83} (1999) 476;\\
 I.~Chepelev and R.~Roiban,``Renormalization of Quantum field Theories on
Noncommutative $R^d$. I: Scalars'', hep-th/9911098, JHEP {\bf 05} (2000)
037;\\ 
B. A. Campbell and K. Kaminsky, ``Noncommutative Field Theory and
Spontaneous Symmetry Breaking'', Nucl. Phys. {\bf B 581} (2000) 240,
hep-th/0003137;\\
A. Matusis, L Ausskind and  N. Toumbas, ``The IR/UV Connection in 
Noncommutative Gauge Theories'', hep-th/0002075.
\item W. H. Huang,  ``Two-Loop Effective Potential in Noncommutative scalar
field theory'',  hep-th/0009067.
\item S. Nam,  ``Casimir Force in Compact Noncommutative Extra Dimensions
and Radius Stabilization'', JHEP {\bf 10} (2000) 044, hep-th/0008083.
\item J. Gomis, T. Mehen and M.B. Wise, ``Quantum Field Theories with
Compact Noncommutative Extra Dimensions'', JHEP {\bf 08} (2000) 029.
 hep-th/0006160. 
\item W. H. Huang,  ``Casimir Effect on the Radius Stabilization of the 
Noncommutative Torus'',  hep-th/0010160.
\item S. J. Gates, H. T. Grisaru, M. Rocek and W. Siegel,  ``{\it
Superspace: One Thousand and One lectures in Supersymmetry}'',
Benjamin/Cummings, 1983;\\
P. West, ``{\it Introduction to the Supersymmetry and Supergavity}'', World
Scientific Publishing, 1986.
\item B. A. Ovrut and J. Wess, Phys Rev. {\bf D 25} (1982) 409;\\
A. Sen,  Phys Rev. {\bf D 31} (1985) 2100;\\
W. H. Huang, Phys. Lett. {\bf B 179} (1986) 92.
\item A. Das and M. Kaku, Phys Rev. {\bf D 18} (1978) 4540;\\
L. Girardello, M. T. Grisaru and P. Salomonson, Nucl. Phys. {\bf B 178}
(1981) 331.
\item W. Fischler, E. Gorbatov, A. Kashani-Poor, S. Paban, P. Pouliot and
J. Gomis, ``Evidence for Winding States in Noncommutative Quantum Field
Theory'',  hep-th/0002067, JHEP {\bf 05} (2000) 024. 
\item J. Schwinger, Phys. Rev. {\bf 82} (1951) 664;\\
        L. Parker, in {\it Recent Development in Gravitation}, edited by S.
Deser         and M. Levey (Plenum, New York, 1977 );\\
        W. H. Huang, Ann. Phys. (N.Y.) {\bf 254} (1997) 69; Phys. Rev. D
{\bf 58} (1998) 084007.
\item  J. Gomis and T. Mehen, ``Spacetime  Noncommutative Field Theories
and Unitarity'', JHEP {\bf 08} (2000) 029 hep-th/0005129. 
\end{enumerate}
\end{document}